# PARAMAGNETIC MOLECULE INDUCED STRONG ANTIFERROMAGNETIC EXCHANGE COUPLING ON A MAGNETIC TUNNEL JUNCTION BASED MOLECULAR SPINTRONICS DEVICE

PAWAN TYAGI[1,3*] COLLIN BAKER[1], CHRISTOPHER D'ANGELO[2]

*University of the District of Columbia, Department of Mechanical Engineering, 4200 Connecticut Avenue NW Washington DC-20008, USA[1]*

Email*: ptyagi@udc.edu

*University of the District of Columbia, Department of Mathematics and Statistics, 4200 Connecticut Avenue NW Washington DC-20008, USA[2]*

*University of Kentucky, Chemical and Materials Engineering Department, 177 F Paul Anderson Hall, Lexington, KY-40506, USA[3]*

This paper reports our Monte Carlo (MC) studies aiming to explain the experimentally observed paramagnetic molecule induced antiferromagnetic coupling between the ferromagnetic (FM) electrodes. Recently developed magnetic tunnel junction based molecular spintronics devices (MTJMSDs), which were prepared by chemically bonding the paramagnetic molecules between the FM electrodes along the exposed side edges of magnetic tunnel junctions, exhibited molecule induced strong antiferromagnetic coupling. Our MC studies focused on the atomic model analogous to the MTJMSD and studied the effect of molecule's magnetic couplings with the two FM electrodes. Simulations show that when a molecule established ferromagnetic coupling with one electrode and antiferromagnetic coupling with the other electrode then theoretical results effectively explained the experimental findings. MC and experimental studies suggests that the strength of exchange coupling between molecule and FM electrode should be ≥50% of the interatomic exchange coupling strength of the FM electrodes.

**1. Introduction:** Molecular spintronics devices (MSDs) have attracted worldwide attention due to their potential to revolutionize logic and memory devices [1, 2]. A typical MSD is comprised of two ferromagnetic (FM) electrodes- coupled by molecular channels [3, 4]. Molecular channels with a net spin state are the basis of a large number of intriguing studies [5], which were either observed experimentally[6, 7] or were calculated theoretically [2]. Porphyrins [8], single molecular magnets [2], and magnetic molecular clusters[9] possess a net spin state and can be synthetically tailored to be employed in a MSD. Single molecular magnet-based MSDs have been widely discussed as the practical architecture for quantum computation [1]. Moreover, paramagnetic molecules strongly coupled to FM electrodes are expected to yield a novel class of magnetic metamaterials and novel device forms. We have recently discussed magnetic tunnel junction (MTJ based MSDs, referred as MTJMSD in this paper, as the most promising, practical, and versatile approach to harness molecule as the device element [3, 10]. This approach necessitates the chemical bonding the molecular

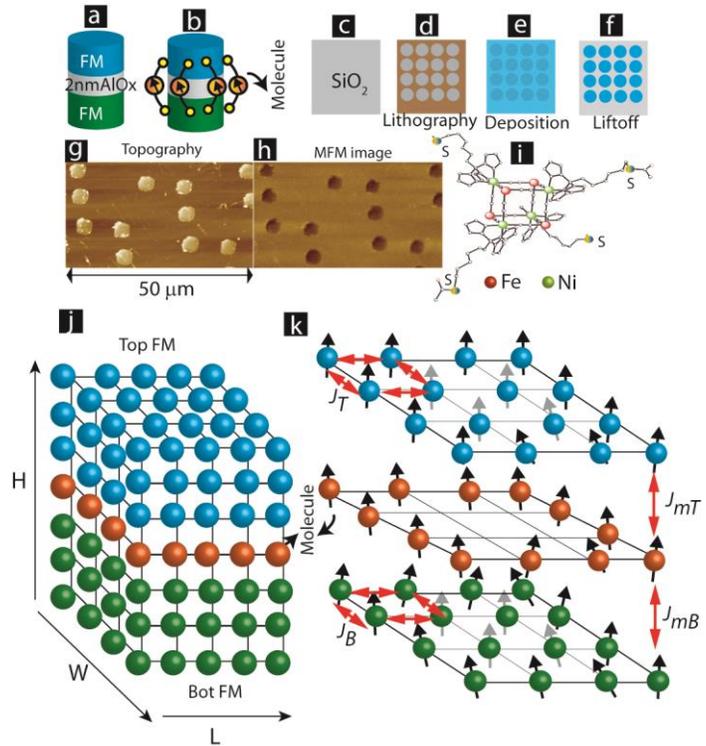

Fig. 1 (a) AMTJ becomes (b) a MTJMSD after hosting paramagnetic molecules. For MTJMSD fabrication a (c) $SiO_2$ covered Si is subjected to (d) photolithography to produce cylindrical cavities for the (e) MTJ deposition followed by (f) liftoff. (g) Topography and (h) MFM image from an actual array of MTJ cylinders. (i) a paramagnetic organometallic molecule with alkane tethers was utilized to transform MTJ into MTJMSD. (j) Analogous atomic model representing MTJMSD in the MC simulations. (k) Top and bottom FMs are only coupled by the molecular coupling along the edges only. $J_T$ and $J_B$ are the nearest neighbor exchange couplings for the top and bottom FM electrodes, respectively.

channels on the FM electrodes of a prefabricated exposed edge MTJ (Fig. 1a) to develop novel MTJMSDs (Fig. 1b). For the first time MTJMSD approach exhibited direct evidence of molecular coupling on the magnetic properties of a MTJ at room temperature. This paper focuses on MC simulations explaining the experimentally observed paramagnetic molecule induced magnetic changes on a prefabricated MTJ. Our MC simulations, which are performed on a theoretical model analogous to MTJMSD, investigated the impact of the nature and magnitude of molecular coupling with the FM electrodes, thermal energy ($kT$), and MTJMSDs sizes.

**2. Experimental details and simulation methodology:**

MTJMSD fabrication protocol involves a flat insulated substrate (Fig. 1c) with microscopic cavities in the photoresist (Fig. 1d) to do sequential depositions of FM electrodes, and ultrathin insulator (Fig. 1e). This yields several thousand MTJs with exposed sides after the liftoff (Fig. 1f). These MTJs with exposed side edges can be structurally (Fig. 1g) and magnetically (Fig. 1h) characterized before introducing paramagnetic molecular device elements; magnetic force microscopy (MFM) image of a cluster of bare MTJs is shown in figure 1h. In this study we used organometallic molecular clusters (OMC). The OMC molecules utilized in this study were synthesized by the Holmes group [9, 11]. In an OMC, the $Fe^{III}$ and $Ni^{II}$ centers positioned in alternate corners of a box and are linked via cyanides (Fig. 1i). Specific details about the thin film depositions [4], MTJ fabrication [12-14]. molecule attachment protocol[4] and OMCs[9, 11, 15] (Fig. 1i) have been published elsewhere. The experimental magnetic studies before and after attaching OMCs- demonstrated unprecedented changes in the magnetic properties of a MTJ [13]. These studies produced strong evidence that molecules are much more than a simple spin or charge carrier. OMCs produced unprecedented, strong antiferromagnetic coupling for the MTJ with Ta (5 nm)/Co (3-5 nm)/NiFe (5-7 nm)/AlOx (2 nm)/NiFe (10 nm) configuration. In this MTJ tantalum (Ta) served as the seed layer. Cobalt (Co) and NiFe(81% Ni/19% Fe) were deposited as the bottom FM electrode followed by the 2 nm alumina (AlOx) tunnel barrier and NiFe top FM electrode.

In order to understand the mechanism behind OMC induced strong coupling we have conducted MC simulations on an analogous MTJMSD system designed in the Ising model framework (Fig.1j). Our previous attempt to explain MTJMSD magnetic properties with non-vector spin and 2D Ising model fall short [14]; to overcome the limitation of previous work we conducted MC study with the actual MTJMSD model and used vector form of the spin. To represent the molecules on the edges, (Fig. 1k), a plane containing atoms along the sides and with empty interior was introduced between the two FM electrodes; FM electrodes are represented by the Ising model. The inter-FM electrode magnetic coupling is only occurring via the molecules (Fig. 1k). However, inter- FM electrode coupling via the empty space is considered to be zero. Using this MC model (Fig. 1j) we performed MC simulations by varying molecular coupling strength with the top FM ($J_{mT}$) and bottom FM ($J_{mB}$) electrodes, $kT$ and MTJMSD dimensions (Fig. 1k). To vary the dimension of a MTJMSD we varied the Height (H), width (W), and Length (L) and overall device dimension is represented by H x W x L (Fig. 1j). Molecular plane is inserted along the $(H-1)/2^{th}$ plane, i.e. the center plane along the H axis of a MTJMSD (Fig. 1j). To achieve the equilibrium state of a MTJMSD under the influence of molecule induced coupling we minimized the system energy as mentioned in eq. 1.

$$E = -J_T(\sum_{i \in T}\vec{S}_i \vec{S}_{i+1}) - J_B(\sum_{i \in B}\vec{S}_i \vec{S}_{i+1}) - J_{mT}(\sum_{i \in T, i+1 \in mol}\vec{S}_i \vec{S}_{i+1}) - J_{mB}(\sum_{i-1 \in mol, i \in B}\vec{S}_{i-1} \vec{S}_i) \qquad (Eq.1)$$

Where, $S$ represents the spin of individual atoms of FM electrodes and molecule in the form of a 3D vectors. In the eq. 1, $J_T$, and $J_B$, are the Heisenberg exchange coupling strengths for the FM electrodes on the top and bottom FM electrodes (Fig. 1k). Our MC studies utilized a continuous model[16] which allowed spin vectors to settle in any direction according to the equilibrium energy governed by eq. 1. For all MC simulations the boundary condition were selected in such a way that the spin of atoms beyond boundary atom of the MTJMSD model (Fig. 1j) were zero.[16] After choosing appropriate values for the Heisenberg exchange coupling coefficients, $kT$, and random spin states, a Markov process was set up to generate a new state. Under the Metropolis algorithm, the spin vector direction of a randomly selected site was changed to produce a new state; energy for the new and old configuration was calculated using eq.1. New states were accepted if the difference between the final and new energy ($\Delta E$) was

$\Delta E<0$ or $exp(-\Delta E/kT) \geq r$.

Where $r$ is a uniformly distributed random variable whose magnitude range from 0 to 1. To achieve a stable low energy state, every MC simulation was run 10 to 100 million steps, depending upon MTJMSD dimensions. After this MC simulations, further runs were performed to generate an average magnitude of observables; two subsequent recordings for any observables were collected at the time interval comparable to autocorrelation time [16]. The units of total energy $E$ and exchange coupling parameters is same as of $kT$. To keep discussion generic, the exchange coupling parameters and $kT$ are referred as the unitless parameters throughout this study. Overall magnetic moment of the MTJMSD is the sum of magnetic moment of the two FM electrodes and the magnetic moment of the molecules.

**3. Results and discussion:** This paper focuses on MC simulations that provide insights about the experimental observations of paramagnetic molecule induced strong antiferromagnetic coupling between the FM electrodes of a MTJ. The following section discusses the experimental results and corresponding MC simulations to provide mechanistic insights about MTJMSDs.

**3.1. Experimental study of MTJMSD:** A MTJ with Ta (5 nm)/Co (3-5 nm)/NiFe (5-7 nm)/AlOx (2 nm)/NiFe (10 nm) configuration demonstrated a dramatic change in its properties after interacting with the OMCs (Fig. 1i). Exclusive studies revealed that OMC bulk possessed S=6 in state around 1K [11]. This spin state decreased to S=3 as temperature increased to 60 K. We were unable to estimate the spin state of those OMCs which got integrated in a MTJMSD. To simplify our MC studies we only considered an S=1 spin state for the molecules throughout. Alkane tethers are expected to serve as the perfect spin channel, as compared to ~ 2 nm AlOx tunnel barrier, with low spin orbit and hyperfine splitting to ensure high spin coherence length and time [17]. Hence a sufficient population of OMCs can serve as the highly efficient spin channels producing strong coupling.

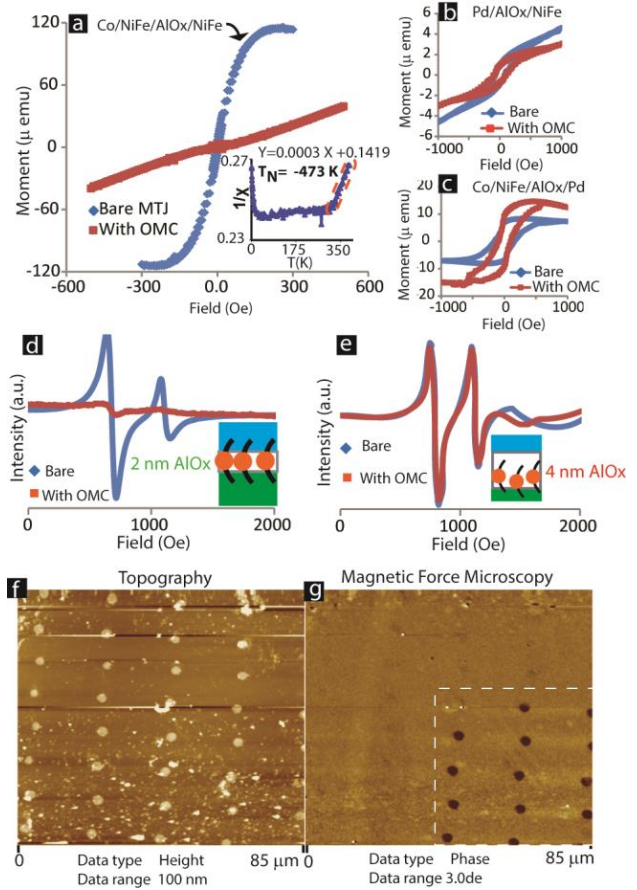

The SQUID magnetometer studies, performed, with Quantum Design MPMS ® showed a typical hysteresis loop of a Co/NiFe/AlOx/NiFe MTJ transformed into a linear magnetic curve (Fig. 2a) after interacting with the OMCs. The inverse of magnetic susceptibility ($\chi^{-1}$) versus temperature ($T$) study showed that a MTJMSD exhibited a prominent transition around 350 K (inset figure of Fig. 2a). Linear fit to the encircled section suggest that for $\chi^{-1} = 0$ the corresponding $T$ was -473 K. Repetition of the same $\chi^{-1}$ vs. $T$ study on another MTJMSD, produced in a different batch, showed corresponding $T$ for $\chi^{-1} =0$ was -404 K. The negative sign of the $T$ for $\chi^{-1} =0$, also known as Neel temperature [18]. Existence of Neel temperature indicates that OMC induced a net antiferromagnetic coupling between the two FM electrodes of the MTJ. We compared the observed OMC induced Neel temperature with the Curie temperature ($T_c$) of the NiFe FM electrode, which directly bonded with OMCs in a MTJMSD. The ratio of OMC induced magnetic coupling, which is of the order of Neel temperature, and ~800K $T_c$ for the NiFe FM

Fig. 2 (a) Magnetization vs magnetic field study of a Co/NiFe/AlOx/NIFe MTJ before after hosting OMCs to become MTJMSD; inset graph shows plot of $\chi^{-1}$ vs. $T$. Tunnel junction with Pd and (b) top NiFe FM (c) bottom Co/NiFe showing opposite response from OMCs . FMR study of Co/NiFe/AlOx/NiFe MTJ with (d) 2nm AlOx and (e) 4 nm AlOx before and after OMCs interaction. (f) Topography and (g) MFM image of Co/NiFe/AlOx/NiFe MTJ based MTJMSD.

electrode [19], was in 0.5-0.54 range. We concluded that OMC induced antiferromagnetic coupling is of the order of 0.5 times of the interatomic ferromagnetic exchange coupling strengths; we assume that $T_c$ corresponds to the interatomic exchange coupling on FMs.

To substantiate our hypothesis that the nature of magnetic interactions of OMCs are opposite with the two FM electrodes -two different types of tunnel junctions were studied. These two tunnel junctions were designed to contain one of the two FM electrode of the MTJMSD and palladium (Pd) as the another electrode. Interestingly, OMCs decreased the magnetic moment of Pd (10 nm)/AlOx (2 nm)/NiFe(12 nm) tunnel junction (Fig. 2b). On the other hand, OMCs increased the magnetic moment of Co(5 nm) /NiFe (5 nm)/AlOx (2nm)/Pd (10 nm) tunnel junction (Fig. 2c). Assuming that OMCs interaction with the Pd was identical in the two cases the results in figure 2b and 2c suggest that OMCs had antiferromagnetic coupling with the NiFe electrode and ferromagnetic coupling with the Co/NiFe electrode. If our interpretation of these experimental studies is correct then MC simulations must exhibit complementary or confirmatory results providing the connection between MTJMSD low magnetization state and necessity of $J_{mT}$ and $J_{mB}$ have opposite sign.

We surmise that such an unprecedented molecule induced antiferromagnetic coupling should also be visible in the other forms of magnetic studies. We performed ferromagnetic resonance (FMR) studies before and after transforming MTJ (Fig. 1a) into MTJMSD (Fig. 1b). It was observed that intensities of typical optical and acoustic resonance modes from the bare MTJs decreased significantly and in some cases disappeared after attaching OMCs on MTJ with Ta (5 nm)/Co (~5 nm)/NiFe (~5 nm)/AlOx (2 nm)/NiFe (~10 nm) configuration; note this MTJ configuration exhibited OMC induced antiferromagnetic coupling during SQUID magnetometer study (Fig. 2a). We also conducted similar experiments on the MTJs with the 4 nm thick AlOx spacer to make sure that OMCs are not able to bridge the gap; no statistical difference was observed due to OMCs. According to Layadi et al. [20], if antiferromagnetic coupling strength between the two FM electrodes increased beyond a critical limit then magnetization of two FM electrodes align antiparallel to each other; in this event two usual resonance modes disappear and only a single mode appear at a higher magnetic field. More importantly, the intensity of the single mode arising after establishing strong antiferromagnetic coupling will be proportional to the square of ($t_T M_T - t_B M_B$); where $t_T$ and $t_B$ are the thickness of top and bottom FM electrodes, respectively; $M_T$ and $M_B$ are the magnetizations of the top and bottom FM electrodes, respectively. Hence, on an OMC affected MTJ with $t_T \approx t_B$ the resultant single mode will be appearing at a higher magnetic field and will possess significantly less intensity as compared to the two modes observed on a bare MTJ. This theoretical study provides explanation to the disappearance of FMR modes and strongly suggests that OMC produced strong antiferromagnetic coupling between two FM electrodes.

To further substantiate the presence of OMC induced strong antiferromagnetic coupling magnetic force microscopy (MFM) studies were conducted. Veeco Multimode AFM and Co coated magnetic cantilever (Nanoscience). It is noteworthy that MFM imaging is based on measuring the change in long range dipolar forces between a magnetic sample and MFM cantilever. We observed that in most of the scans at MTJMSD coordinates of the topographical images (Fig. 2f) extremely faint or negligible magnetic contrast was observed (Fig. 2g). This study is important in providing evidence that MTJMSD are physically intact. Interestingly, in some MFM scans coexistence of high MFM contrast and negligible MFM contrast was observed. We believe that high contrast MFM is arising from those MTJs which failed to transform into MTJMSD after interacting with the OMCs. On the positive side, such imperfect MTJMSD serve as a good reference to justify the validity of the MFM imaging parameters.

### 3.2 MC study of MTJMSD:

Molecule's couplings with the top and bottom FM electrodes are the two most important parameters in governing the magnetic properties of a MTJMSD (Fig. 1k). We first varied $J_{mT}$ and $J_{mB}$ at fixed $kT$ to investigate which combination of the molecular couplings yields the antiferromagnetic couplings between FM electrodes leading to the experimental observations on MTJMSD (Fig. 2). A 3D graph for 11x10x10 MTJMSD at $kT$=0.1 suggests the M (magnetic moment of the MTJMSD model) was approaching the magnitude of net molecular magnetic moment when $J_{mT}$ and $J_{mB}$ were of opposite signs (Fig. 3a); it does not matter if $J_{mT}$ or $J_{mB}$ is positive or negative. This MC result (Fig. 3a) confirms our interpretation of experimental magnetization data (Fig. 2b-c) that OMC developed ferromagnetic (+) coupling with the Co/NiFe electrode and antiferromagnetic (-) coupling with the NiFe electrode; hence, in order to see the near zero MTJMSD magnetization $J_{mT}$ and $J_{mB}$ must be of opposite sign (Fig. 3a).

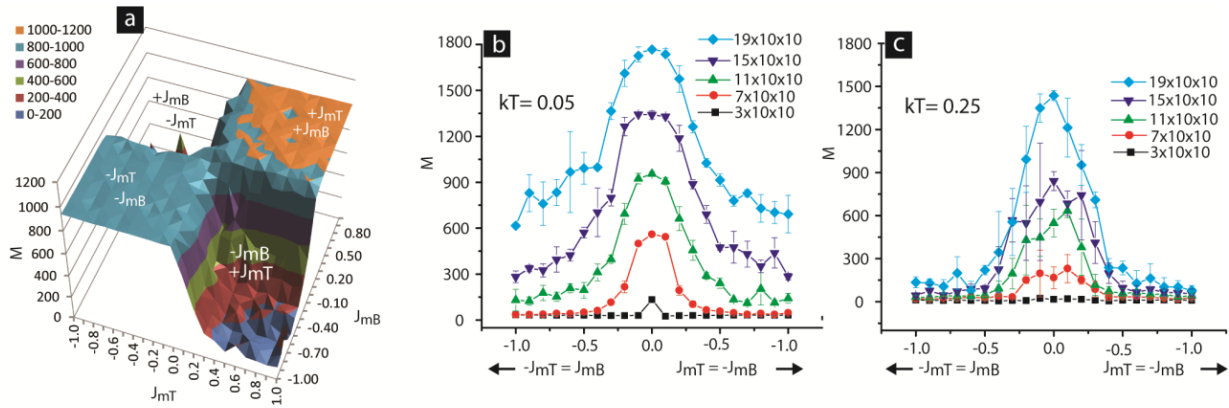

Fig. 3: (a) Effect of molecular coupling with top electrode ($J_{mT}$) and bottom electrode ($J_{mB}$) on overall magnetization of the MTJMSD at $kT = 0.1$ . Effect of equal magnitude and opposite nature $J_{mT}$ and $J_{mB}$ on MSD magnetization for the different size MSDs at (b) $kT = 0.05$ and (c) $kT = 0.25$

Our MC studies also investigated the effect of the magnitude of $J_{mT}$ and $J_{mB}$ and MTJMSD device size. For this study *M* for various MTJMSD sizes was plotted as a function of -equal and opposite values of $J_{mT}$ and $J_{mB}$ , i.e. $J_{mT} = - J_{mB}$ or $-J_{mT} = J_{mB}$ (Fig. 3b-c). We varied the height of MTJMSD with (Hx10x10) size (Fig. 1j) to vary the number of atoms of the FM electrode without changing the number of molecules. For low thermal energy at $kT = 0.05$ it was noticed that strength of $J_{mT}$ and $J_{mB}$ required to bring overall MTJMSD's M close to the number of molecules increased with the device size. For Instance, 3x10x10 and 7x10x10 MTJMSD settled in low *M* state when $|J_{mT}|$ and $|J_{mB}|$ were ~0.1 and ~0.5, respectively. Further increase in device size made it very difficult for $J_{mT}$ and $J_{mB}$ to bring M of MTJMSD close to zero. For low *kT* ordered molecules are responsible for the MTJMSD's M, when both FM electrodes are aligned in the opposite direction. Interestingly increasing $kT$ assisted the MTJMSD to settle in low magnetization state; at $kT = 0.25$ most of MTJMSD sizes settled in near zero magnetization state. However, higher $kT$ also disordered the molecular ordering to transcend MTJMSD, with oppositely aligned FMs, near zero. It is evident that thermal fluctuations can magnify the impact of molecule strength. Also, we noticed that when $J_{mT}$ and $J_{mB}$ approached 0.5 only then MTJMSD's magnetization could settle around zero (Fig. 3c). Interestingly, this MC simulation results support the experimental observations (inset graph of Fig. 2a). Since $kT_c$ of FM electrodes in the MC study is of the order of $J_T = J_B = 1$, hence the ration of $J_{mT}$ or $-J_{mB}$ to $kT_c$ will be of the order of 0.5. Interestingly, the ratio of molecule induced antiferromagnetic coupling, represented by the Neel temperature, and $T_c$ was in 0.5-0.54 range (Fig. 2a). The MC simulation appears to correctly estimate the order of molecule induced exchange coupling strength; however, it is important to note that more accurate estimation will require the inclusion of FM electrode anisotropy, and other forms of couplings via molecules, such as biquadratic coupling, dipolar coupling, etc.

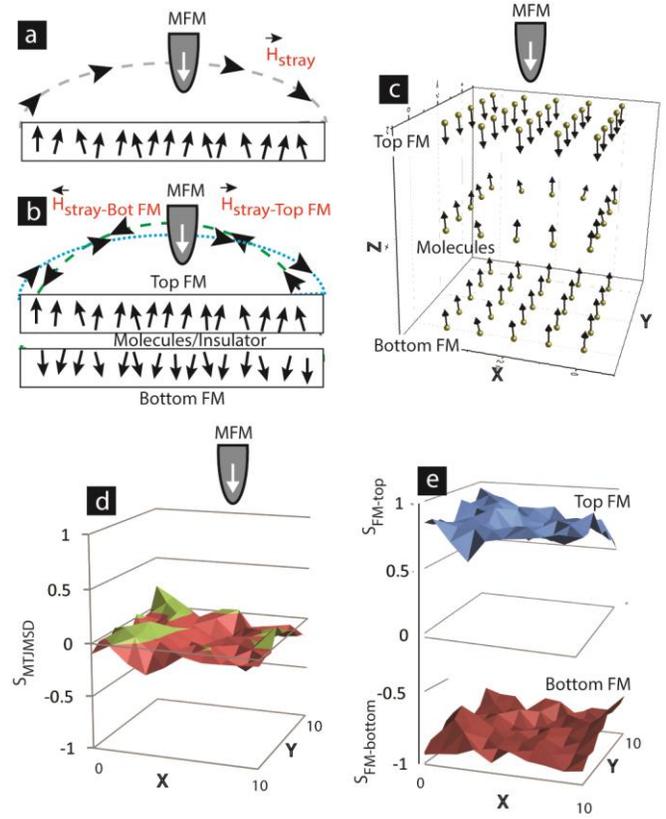

Fig 4. (a) MFM tip monopol experiencing stray magnetic field arising from a ferromagnetic film. (b) A MSD with two oppositely aligned FM electrodes is expected to produce two stray magnetic fields acting in opposite directions. (c) MC simulation produced molecule coupling induced opposite orientation of magnetic moments from the two FM electrodes. Dot product of the MFM probe moment and (d) average MSD magnetic moment settled near zero. However, dot product of the MFM probe moment and (e) top FM electrode and (b) bottom FM electrode of the same MTJMSD was near 1 and -1, respectively.

MC simulations also complement the FMR experimental studies on MTJMSDs (Fig. 2c). Layadi et al. [20] have theoretically described for very strong antiferromagnetic coupling forces two FM electrodes to align in the opposite direction and the resultant FMR spectra from such a system resembles the FMR data obtained from MTJMSD (Fig. 2c). Our MC simulation also showed that MTJMSD with opposite signs of $J_{mT}$ and $J_{mB}$ lead to the opposite alignment of FM electrodes. We also attempted to gain insights about the MFM studies which showed negligible magnetic contrast at the sites of MTJMSDs (Fig. 2g). MFM images are a result of magnetic force ($F$) experienced by the magnetic tip's magnetization ($m$) in the stray field ($H$) generated by the magnetic sample in the ambient of magnetic permittivity ($\mu$) (Fig. 4a). It is given by the following equation.

$$\vec{F} = \mu(\vec{m}.\nabla)\vec{H} \qquad \text{eq. 2}$$

We hypothesized that the oppositely aligned top FM and bottom FM will produce stray magnetic fields in the opposite direction to yield negligible magnetic contrast from the MTJMSD (Fig. 4b). Our MC simulation generated the atomic and molecule site specific magnetization vector profile; an atomic scale 3D vector plot of 3x5x5 MSD device size for $J_{mT}= -J_{mB} =0.5$ and $kT=0.1$ is presented (Fig. 4c). At the first place this 3D view asserts that our MC studies are working on the right model which is analogous to the MTJMSD device (Fig. 1b). We performed similar studies on 11x10x10 MSD size and calculated spatial magnetic moment plot for the MTJMSD and the two FM electrodes (Fig. 4 d-e). For MTJMSD we summed the magnetic moment of atoms of FM electrodes and molecules at each topological site along the height dimension and it turned out be very close the total magnetic moment of the molecules, which is only 6.9% of the total magnetization of FM electrode for the 11x10x10 MTJMSD; 36 molecules per 500 FM atoms. Such a small spatial magnetic moment at each spatial site will produce negligible stray field and magnetic contrast as observed in the experimental MFM image from actual MTJMSD (Fig. 2g). As shown in the schematic stray field, the average magnetization of the oppositely aligned FM electrodes (Fig. 4b) will cancel each other. However, independent measurement of spatial distribution of the average magnetic moment of the top and bottom FM electrode will still be very high as compared to that of overall MTJMSD (Fig. 4e).

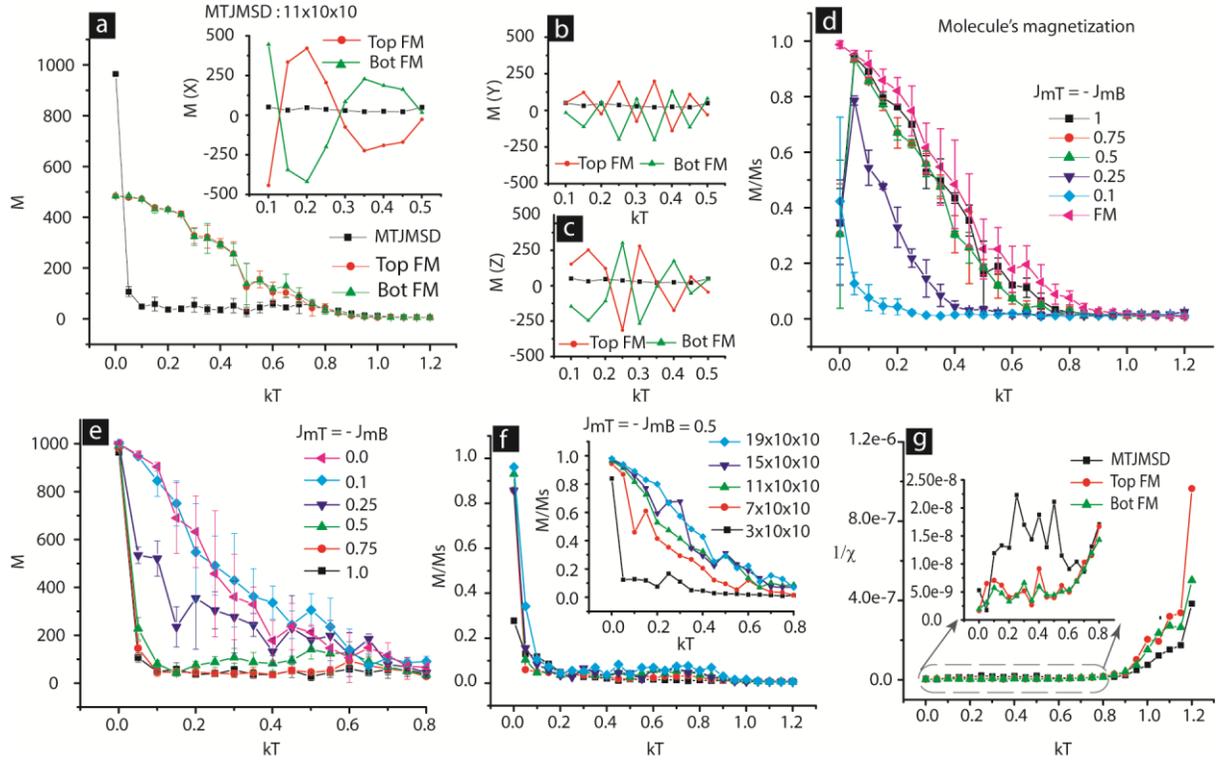

Fig. 5. (a) Magnetic moment (*M*) of 11x10x10 MSD and its FM electrode vs. *kT* for $J_{mT}=-J_{mB}=1$. Inset graph of (a) shows the X component (M(X)), (b) Y component (M(Y)), and (c) z component (M(Z)) of the average magnetic moment vector for the top and bottom FMs. (d) Normalized M of the paramagnetic molecules vs. *kT* graph for equal and varying strength of $J_{mT}$ and $-J_{mB}$. (e) M of 11x10x10 MSD vs. *kT* for equal and varying strength of $J_{mT}$ and $-J_{mB}$. (f) Normalized magnetic moment of varying MTJMSD sizes at $J_{mT}=-J_{mB}=1$; inset graph correspond to $J_{mT}=-J_{mB}=0.5$. (g) $\chi^{-1}$ vs. kT for 11x10x10 MTJ MSD with $J_{mT}=-J_{mB}=1$; inset graph shows the zoomed in version of limited *kT* range.

In addition to $J_{mT}$ and $J_{mB}$ the variation of *kT* produced a pronounced effect on the MTJMSD magnetic state. For $J_{mT} = -J_{mB}$, as *kT* increased the MTJMSD's M, dropped close to the total magnetization of molecules (Fig. 5a); however, for higher *kT* thermal fluctuations forced MTJMSD magnetization to settle close to zero. For $kT$ being of the order of $J_T$ or $J_B$ (i.e ~$kT_c$) thermal fluctuations smeared the effect of $J_{mT}$ and $J_{mB}$. Transition from high magnetization to low M happened before $kT = 0.1$ for the 11x10x10 MTJMSD (Fig. 5a). Interestingly, the magnetization of the two FM electrodes decreased gradually and followed the trend $(1-T/T_c)^\alpha$. The magnitude of exponent $\alpha$ for our $M$ vs. $kT$ graph for FM electrodes was in the 0.4-0.5 range; this magnitude of $\alpha$ is in the close agreement with the prior literature [21] [18]. Deeper insights about the molecule induced exchange coupling comes from the study of vector components of the top and bottom FM electrodes. For instance the X component of the magnetization vector for the top and bottom FM electrodes accounted for the major alignment direction. However, the X component of the top and bottom FMs aligned in the opposite directions (inset graph of Fig. 5a). Similarly, Y and Z components of two FM electrodes were aligned in the opposite direction for the 0.1 to 0.8 $kT$ range (Fig 5b and c). The opposite orientation of the FM electrode vectors directly agrees with the atom specific spatial orientation of the spin vector as shown in figure 4c. We also studied the effect of molecular coupling strength on the magnetization of the molecules (Fig. 5d). It is apparent that molecules were well ordered when molecular coupling strength was ~0.5 or more; for the weaker coupling strengths molecules assumed random spin vectors (Fig. 5d). A MTJMSD experienced difficulty in settling in a low magnetization state when molecular coupling strength was <0.5 (Fig. 5e). This result is in agreement with the study focusing on the variations of the molecular coupling strength in device size (Fig. 2). We also studied the effect of MSD size on the M vs. $kT$ graph. For $J_{mT}= -J_{mB} =1$ all the studied MTJMSD sizes settled in near zero magnetization state (Fig. 5f). However, for $J_{mT}= -J_{mB} =0.5$ only the smaller device sizes tended to settle in the lower magnetization state (inset of Fig. 5f). We also studied $\chi^{-1}$ vs. $kT$ for 11x10x10 MSD size (Fig. 5g). This study suggests that a major transition occurred close to the $kT_c$ (or $kT_c =J_T=J_B$) for the MTJMSD (Fig. 5g). Zooming on the data enclosed in the gray color lines showed that $\chi^{-1}$ of the overall MTJMSD was more than that of FM electrodes, before the $kT_c$ (Inset of Fig. 5g). We believe that this region of $kT$ signifies the dominance of the molecular coupling. However, after the Curie temperature ($J_T=J_B$) $\chi^{-1}$ for the FM electrodes dominated. Presumably $kT$ destroyed the ordering due to $J_T$ and $J_B$ on the FM electrodes. This study suggests that the effect of molecular coupling ($J_{mT}$ and -$J_{mB}$) was functional up to $kT=\sim0.8$.

**4. CONCLUSIONS**

MC simulations were performed to study the effect of magnetic molecule induced exchange coupling on the magnetic properties of the MTJMSD. We considered all the possible permutations and combinations and nature of interactions between a paramagnetic molecule and the two FM electrodes of a MTJMSD to understand the experimental results. Experimentally observed molecule induced strong antiferromagnetic coupling was only possible when a molecule, with a net spin state, established ferromagnetic coupling with one FM electrode and antiferromagnetic coupling with the other FM electrode. Our MC simulations effectively explain the origin of the experimental data obtained from SQUID magnetometer, ferromagnetic resonance, and MFM studies on MTJMSD. The experimentally estimated molecular coupling strength was in agreement with our results of MC simulations. Increasing MTJMSD size was found to weaken the molecular coupling effect. However it is quite possible that we underestimated the impact of molecular coupling on the MTJMSD size. In this study we mainly focused on the Heisenberg type magnetic interaction among nearest neighbors. In reality, molecules are expected to have other modes of couplings such as biquadratic coupling, dipolar coupling, and most importantly paramagnetic molecules are also capable of invoking spin fluctuation assisted coupling between two FM electrodes.[5] One significant caveat about our MC simulation is that it considers FM electrodes to be 100% spin polarized; however, in actual a FM electrode is nearly 40% spin polarized.[22, 23] We surmise that assuming 100% spin polarized FM electrodes is still a good assumption in the context of MTJMSDs. It is because of the fact that OMC induced strong coupling is expected to produce spin filtering leading to highly spin polarized FM electrodes. Molecular channels with small spin –orbit coupling and hyperfine splitting, can ensure high spin coherence as compared to a ~2 nm AlOx insulator with numerous spin scattering defect sites and imperfections. Further experimental and theoretical studies are needed in order to explore the rich physics and novel device forms associated with the MTJMSD approach.

**ACKNOWLDGEMENT**

The MCS study was in part supported by National Science Foundation-Research Initiation Award (Contract # HRD-1238802), Department of Energy/ National Nuclear Security Agency subaward (Subaward No. 0007701-1000043016). We thankfully acknowledge the Air Force Office of Sponsored Research (Award #FA9550-13-1-0152) for facilitating study of ferromagnetic electrode stability. Pawan Tyagi thanks Dr. Bruce Hinds and Department of Chemical and Materials engineering at University of Kentucky for facilitating experimental work on tunnel junction based molecular devices during PhD. Molecules for molecular device fabrication were produced Dr. Stephen Holmes's group. Any opinions, findings, and conclusions expressed in this material are those of the author(s) and do not necessarily reflect the views of any funding agency and corresponding author's past and present affiliations.

**References:**

[1]   Coronado E and Epsetin A J 2009 Molecular spintronics and quantum computing *J. Mater. Chem.* **19** 1670-1

[2]   Bogani L and Wernsdorfer W 2008 Molecular spintronics using single-molecule magnets *Nat. Mater.* **7** 179-86

[3]   Tyagi P, Friebe E and Baker C 2015 ADVANTAGES OF PREFABRICATED TUNNEL JUNCTION BASED MOLECULAR SPINTRONICS DEVICES *NANO* **10** 1530002

[4]   Tyagi P, Li D F, Holmes S M and Hinds B J 2007 Molecular electrodes at the exposed edge of metal/insulator/metal trilayer structures *J. Am. Chem. Soc.* **129** 4929-38

[5]   Pasupathy A N, Bialczak R C, Martinek J, Grose J E, Donev L A K, McEuen P L and Ralph D C 2004 The Kondo effect in the presence of ferromagnetism *Science* **306** 86-9

[6]   Heersche H B, de Groot Z, Folk J A, van der Zant H S J, Romeike C, Wegewijs M R, Zobbi L, Barreca D, Tondello E and Cornia A 2006 Electron transport through single Mn-12 molecular magnets *Phys. Rev. Lett.* **96** 206801

[7] Liang W J, Shores M P, Bockrath M, Long J R and Park H 2002 Kondo resonance in a single-molecule transistor *Nature* **417** 725-9

[8] Jurow M, Schuckman A E, Batteas J D and Drain C M 2010 Porphyrins as molecular electronic components of functional devices *Coord. Chem. Rev.* **254** 2297-310

[9] Li D F, Parkin S, Wang G B, Yee G T, Clerac R, Wernsdorfer W and Holmes S M 2006 An S=6 cyanide-bridged octanuclear (Fe4Ni4II)-Ni-III complex that exhibits slow relaxation of the magnetization *J. Am. Chem. Soc.* **128** 4214-5

[10] Tyagi P 2011 Multilayer edge molecular electronics devices: a review *J. Mater. Chem.* **21** 4733-42

[11] Li D F, Ruschman C, Clerac R and Holmes S M 2006 Ancillary Ligand Functionalization of Cyanide-Bridged S = 6 FeIII4NiII4 Complexes for Molecule-Based Electronics *Inorg. Chem.* **45** 7569

[12] Tyagi P and Hinds B J 2010 Mechanism of Ultrathin Tunnel Barrier Failure Due to Mechanical Stress Induced Nano-Sized Hillocks and Voids *J. Vac. Sci. Technol. B* **28** 517-21

[13] Tyagi P 2013 Molecule Induced Strong Coupling between Ferromagnetic Electrodes of a Molecular Spintronics Device *Mater. Sci. Foroum* **736** 32-54

[14] Tyagi P, D'Angelo C and Baker C 2015 MONTE CARLO AND EXPERIMENTAL MAGNETIC STUDIES OF MOLECULAR SPINTRONICS DEVICES *NANO* **10** 1550056

[15] Park K and M. H S 2006 Exchange coupling and contribution of induced orbital angular momentum of low-spin Fe3+ ions to magnetic anisotropy in cyanide-bridged Fe2M2 molecular magnets: Spin-polarized density-functional calculations *Phys. Rev. B* **74** 224440

[16] Newman M E and Barkema G T 1999 *Monte Carlo Methods in Statistical Physics* (Oxford: Clarendon Press)

[17] Ouyang M and Awschalom D D 2003 Coherent spin transfer between molecularly bridged quantum dots *Science* **301** 1074-8

[18] Coey J M 2010 *Magnetism and magnetic materials*: Cambridge University Press)

[19]     Johnson M 2004 *Magnetoelectronics*: Academic Press)

[20]     Layadi A 1998 Ferromagnetic resonance modes in coupled layers with cubic magnetocrystalline anisotropy *J. App. Phys.* **83** 3738-43

[21]     Bennett L, McMichael R, Swartzendruber L, Shull R and Watson R 1992 Monte Carlo and mean-field calculations of the magnetocaloric effect of ferromagnetically interacting clusters *J. Mag. Mag. Mat.* **104** 1094-5

[22]     Miao G X, Munzenberg M and Moodera J S 2011 Tunneling path toward spintronics *Rep. Prog. Phys.* **74** 036501

[23]     Moodera J S, Nassar J and Mathon G 1999 Spin-tunneling in ferromagnetic junctions *Ann. Rev. Mater. Sci.* **29** 381-432